# Melting curves of Al,Cu,U and Fe metals utilizing the Lindemann-Gilvarry criterion and parameterization of the equations of state

Joseph Gal
Ilse Katz Institute for Nanoscale Science and Technology ,
Ben-Gurion University of the Negev, Beer Sheva ,84105 Israel

## Abstract

The prediction of the melting curve of metals by extrapolation to high pressures and temperatures based on the Lindemann-Gilvarry criterion (LG) assuming harmonic Debye solid is presented. The LG formulation uses the bulk modulus B and its pressure derivative B' as fit parameters deduced directly from the equation of state, however, the results are not unique and strongly depends on the chosen equation of state (EOS). By introducing a constraint that the bulk moduli parameters B and B' must simultaneously obey the Lindemann-Gilvarry criterion (LG) and the EOS, consistent bulk moduli are derived. The cold pressure $P_c$ and the cold melting curve are obtained by introducing an effective Grüneisen parameter ($\gamma_{eff}$) to the LG approximated equation together with the above constraint. It is claimed that isochoric condition exists in diamond anvil cells (DAC), thus upon raising the temperature and approaching the melt constant volume is maintained. Isochoric condition in the DAC means that the developed thermal pressure ($P_{th}$) should be accounted in the LG formulation. Therefor, the actual pressure ($P_c+P_{th}$) sensed by the sample confined in the DAC should be inserted to the LG melting formula. This brings along the demand that the shock waves Hugoniot melting data should serve as anchor for deriving the correct melting curves of metals where the Grüneisen parameter $\gamma$ at ambient conditions ($\gamma_o$) is directly determined. The melting curves up to ultra high pressures of Al, Cu, U and Fe metals are presented and discussed. In this manner, special attention is given to ε-Fe as isobaric condition in the DAC has been claimed. Utilizing the present approach we obtain the melting temperature of iron in the earth inner core boundary (ICB, 330GPa) is 5900±100K.

1. Introduction

The determination of the pressure dependent melting temperatures of solids has drawn the attention of the scientific community for many years.
In the present contribution we apply Lindemamm's criterion for prediction of melting curves, though this criterion is not a theoretical model based on first principles but a phenomenological approach to the behavior of solids. We adopt and trust the Lindemann criterion improved by Gilvarry, known as Lindemann-Gilvarry (LG) criterion [1]. Prediction of the melting point at high pressures and temperatures for metals utilizing the LG criterion needs the Grüneisen parameter $\gamma$. The procedure utilizing the LG criterion together with Grüneisen parameter $\gamma$ according to the Slater model [2] often does not fit the experimental melting results. We therefore propose a different fitting procedure which takes into account simultaneously the LG criterion together with the equation of state (EOS) in the P-V space. In this procedure the shock waves experimental melting data serve as anchor to determine the actual melting curve measured in a diamond anvil cell (DAC). The LG formulation uses the bulk modulus B and its pressure derivative B' as fit parameters deduced directly from the EOS, however, the results are not unique. Numerous EOS are available most of them need two free parameters; the bulk moduli B and B' which are deduced from the P-V room temperature isotherm and are assigned $B_o$ and $B_o$'. However, the fitting of the experimental data in the P-V space strongly depends on the chosen equation of state. The reported values of $B_o$ and $B_o$' span up to ~ 50% (see table I) and the question remain which bulk moduli should be addressed.

**Table I**: Elastic bulk modulus $B_o$ in GPa and its pressure derivative $B_o$' derived by the present combined approach compared to those reported in the literature.

|          | $B_o$   | $B_o$' | $B_o$ span      | $B_o$'span   |
|----------|---------|--------|-----------------|--------------|
| Aluminum | 73(1)   | 4.45   | 72 -77 [5,10,13]| 4.0 – 4.54   |
| Copper   | 142(2)  | 4.9    | 133-142 [18,35,36] | 4.54 – 5.0 |
| Iron     | 163(1)  | 5.55   | 163-193 [10,11] | 4.2 - 5.38   |
| Uranium  | 136(2)  | 3.8    | 104-147 [12,18] | 3.8- 6       |

The bulk moduli are of basic importance for the prediction of melting curves at high pressures and temperatures for materials utilizing the LG criterion. The melting curves of metals have been extensively studied by diamond anvil cells and shock wave (SW) experiments. Tremendous advances have been reached in developing high pressure cells using laser heating techniques and the determination of melting points up to very high pressures and temperatures are available. Noteworthy are DAC measurements of Tateno et al. [3] reaching about ~370GPa. Above 370GPa the only available data come from first principle theoretical calculations. Both the DAC and SW techniques have their problems as discrepancies between the experimental melting data exist indicating the difficult interpretation of these experiments. Many possible explanations of discrepancies between the DAC data and SW results have been set forward but none of them are widely accepted. One source of uncertainty is the question of processes taking place in the sample while being heated under pressure. One possibility is that heating takes place under isochoric conditions in which case thermal pressure should be added. In the other extreme the process is isobaric in which case volume change exist. These two extreme cases determine a range of uncertainties in the interpretation. In the present contribution we claim that isochoric condition exist in the DAC experiments and the thermal pressure should be accounted for correction of the LG cold melting curve. Thus the discrepancies between the DAC and the SW measurements are settled. Moreover, the SW Hugoniot melting data should serve as an anchor to determine the actual melting curve. As will be shown below these assumptions are confirmed in the cases of Aluminum copper uranium. However, in case of ε-iron the claim needs a special attention and a further discussion which is given bellow.

## 2. Theory and extrapolation method

According to Lindenmann's criterion the melting temperature $T_m$ is related to the Debye temperature $\Theta_D$ as follows:

$$T_m = C\, V^{3/2}\, \Theta_D^2 \qquad (1)$$

Where V is the volume and C is a constant to be derived for each specific metal.

In the Debye model the Grüneisen parameter $\gamma$ is defined by

$$\gamma = \partial \ln \Theta_D / \partial \ln V \qquad (2)$$

As shown by Anderson and Isaak [4] combining (1) and (2) and inserting $V_o/V = \rho/\rho_o$, and integrating one gets the form of LG criterion of the melting temperature $T_m$ :

$$T_m(\rho) = T_{mo} \exp \left\{ \int_{\rho_o}^{\rho} [2\gamma - 2/3] \, d\rho/\rho \right\} \qquad (3)$$

Where $\rho_o$ is a reference density, $\rho$ is the density at the melt and $T_{mo}$ is the melting temperature at the reference density.

Integrating (3) assuming that

$$\gamma = \gamma_o (\rho_o/\rho)^q \qquad (4)$$

Taking $q=1$ one gets:

$$T_m(\rho) = T_{mo} (\rho_o/\rho)^{2/3} \exp[2\gamma_o (1 - \rho_o/\rho)] \qquad (5)$$

where $\gamma_o$ is Grüneisen parameter at ambient conditions.

Equation (5) states that if $\rho(P)$, $T_{mo}$ and $\gamma_o$ are known the melting curve $T_m(P)$ can simply be determined assuming that the relation between P and $\rho$ is known. It is well accepted that the pressure is given by:

$$P(V,T) = P_C + \gamma_{lattice} C_{v\,lattice} \rho [T - T_o + E_o/C_{v\,lattice}] + \tfrac{1}{4} \rho_o \gamma_e \beta_o (\rho/\rho_o)^{1/2} T^2 \qquad (6)$$

See for example Altshuler ref.[7] and Kormer ref.[8]. Here Pc is the cold pressure, $C_v$ is the lattice specific heat above $T_o$, $T_o$ is the ambient temperature. $C_{v\,lattice}$ is taken as constant (usually at room temperature, following the approximation of Altshuler et al. [7] ), $E_o$ is the lattice thermal energy at $T_o$ and $\gamma_{lattic}$ is the lattice Grüneisen parameter. $\gamma_e$ is electronic Grüneisen parameter and $\beta_o$ is the electronic specific heat coefficient. It is customary to analyze EOS and melting experiments in terms of room temperature isotherms using Murnaghan [4], Vinet [5] and Birch-Murnaghan [6] EOS for deriving $P_c$. The parameters of these equations are the ambient condition bulk modulus $B_o$ and its pressure derivative $B_o'$. In DAC experiments the volume of the tested material close to the melt must be known and is essential to apply the LG theory. Under the assumption of isochoric conditions one can directly use the measured volume in compression at room temperature. It can be also measured insitu while heating the compressed sample. In most experiments the material is compressed at room temperature and then heated to the melting point. Such experiments usually present the measured pressure ( Ruby's line shifts) at ambient temperature. These experiments ignore the thermal contribution (actual pressure) and the results are usually presented as the melting temperature vs. the cold pressure at ambient temperature, known as the cold melting curve.

The relation between $P_C$ and the density $\rho(P)$ for the room temperature isotherms, are given by Murnaghan (MUR) [7], Vinet (VIN) [8], Birch-Murnaghan(BM) [9] equations of state (EOS). In practice BM and VIN EOS better fit the experimental data in the P-V plane and are commonly preferred.

$$P_c = 3B_o (\rho/\rho_o)^{-2/3} [1-(\rho_o/\rho)^{1/3}] \exp\{3/2( B'-1)[1 -(\rho_o/\rho)^{1/3}]\} \quad \text{VIN} \quad (7)$$

and

$$P_c = 3/2\, B_o [(\rho/\rho_o)^{7/3}-(\rho/\rho_o)^{5/3}] [1+3/4(B'-4)\{ (\rho/\rho_o)^{2/3}-1\}] \quad \text{BM} \quad (8)$$

Where $\rho$ is density and $B = -V (\partial P/\partial V)$ is the definition of the bulk modulus and B' is the pressure derivative of the bulk modulus ($B' = \partial B/\partial P$). B and B' are fit parameters of the room temperature isotherm assigned as $B_o$ and $B_o'$. It is well known that the best fit solutions are not unique and strongly depend on the chosen EOS (eq.7,8). This is the reason why diverse results are obtained by different authors as shown in Table I. For example, for ε-iron estimates of $B_o$ span between 160 and 190 GPa [10,11] or $B_o$ of α-Uranium varies between 104 and 138GPa [12] (see there Table I, and the present figures 1-4(c)). Thus, it make sense to introduce a different procedure in order to improve the fitting of the data in the P-V and in the P-T planes using the above equations of state.

We therefore propose an additional constraint for deriving $B_o$ and $B_o'$, namely $B_o, B_o'$ must simultaneously fit the EOS (eqs.7 or 8) and the pressure calculated for each melting point using Lindemann-Gilvarry criterion eq. 5.

In most DAC melting measurements the results are displayed as 300K pressure vs. melting temperature. For each melting point the volume close the melt can be evaluated using the LG criterion. In this way a 300K pressure vs. melting temperature curve can be obtained and optimized against the experimental data using $B_o, B_o'$ and $\gamma$ as parameters. The final optimization $B_o, B_o'$ also includes the EOS measurements data. In this manner $B_o, B_o'$ are consistent with LG $T_m(\rho)$ and can best fit all pressures experimental data with one free parameter $\gamma_{eff}$ replacing $\gamma_o$ in eq.5. Extrapolation to very high pressures and temperatures with the above obtained constants ($B_o, B_o', \gamma_{eff}$) predict the measured melting curve forming the cold pressure $P_c$ and the cold melting curve. It is proposed that the derived $B_o, B_o'$ are those that should be addressed when calculating the LG melting curve.

Our basic assumption is that in an ideal DAC under each applied pressure, starting from ambient pressure and temperature, as raising the temperature and approaching the melt, the metal should stay in isochoric condition. Isochoric condition in the DAC means that the thermal pressure $P_{th}$ and the melting temperature Tm increase upon heating the sample. Thus actual pressure eq.6 P(V,T) should be used for deriving the extrapolated melting curve. The present procedure applies to materials where no phase transition occurs upon raising either the pressure or the temperature (see discussion).

We thus propose the following four steps procedure to determine the correct melting curve (the combined approach):

1. Utilizing Lindemann-Gilvarry criterion (eq.5) with $\gamma_{eff}$ as free parameter and optimizing $B_o$ and $B_o'$ by choosing the appropriate EOS (out of eqs. 7 or 8) which best fit simultaneously the experimental P-V data (isotherm 300K) and the experimental melting P-T data. Thus obtaining $P_c$ forming the cold melting curve. In LG eq.5 $Tm_o$ and $V_o$ are the melting temperature and volume at ambient pressure.

2. Adding the calculated thermal pressure $P_{th}$ to $P_c$ obtaining the LG melting curve accounting for the actual pressure (isochoric condition) sensed by the investigated sample. Demanding that the thermally corrected melting curve will include the shock wave melting data as anchors. The Grüneisen parameter $\gamma_o$ is derived accordingly.

3. Plotting the volume compression V/Vo vs. the thermally corrected melting temperatures obtained in 2.

4. Extrapolating the derived thermally corrected melting curve to high pressures and temperatures.

Claiming isochoric condition in the DAC upon heating the sample confined in the cell thermal pressure develops associated with increase of the melting temperature. The calculated thermal pressure ($Po_{th}$) and the melting point $Tm_o$' at ambient pressure are derived by calculating $Po_{th}$ according to eq.6 and adjusting $\gamma_o$ to match the shock wave data forming the melting curve. To clarify, $Po_{th}$ is the pressure shift from zero pressure and $Tm_o$' is the melting temperature at $Po_{th}$.

In the following several examples are shown and discussed:

## 1. Aluminum

The first examination of our combined approach is aluminum metal. Aluminum at ambient pressure exhibit the fcc structure. It is claimed that by raising the pressure phase transitions occur (fcc => hcp => bcc). However, within the resolution of the x-ray diffraction (XRD) these transitions are not observed indicating that the pressure dependent phase transformations associated with volume changes are indeed very small at least up to 300GPa. Applying step 1. the experimental melting points and the experimental data in the P-V space are best fitted simultaneously with the LG criterion (eq.5) with Vinet EOS. The elastic parameters obtained are $B_o$=73(0.5)GPa and $B_o$'=4.45(0.1) forming $P_c$. The cold melting curve which best fit the experimental melting points yield $\gamma_{eff}$= 2.45(1) where $Tm_o$=995K is the melting point measured in the DAC at 0.75GPa according to Errandonea [14]. The cold melting curve with the above parameters is shown by blue solid line in Fig.1a. The experimental melting data measured by DAC are taken from Hanstrum and Lazor [13] and Errandonea [14] (red O's and X) and Ross et al. [15] (purple stars). The shock waves data marked red diamonds are taken from Ross et al. [15]. The actual melting curve of Aluminum metal (red line Fig.1a) is obtained according the procedure 1-2 where thermal pressure $P_{th}$ was calculated according to Altshuler et al.[8] (there in Table 4). Assuming isochoric condition in the DAC (constant volume) upon raising the temperature up to the melting point the calculated thermal pressure shift is 4GPa ($Po_{th}$) yielding melting at 1150(10)K ($Tm_o$'). Adding the thermal pressure $P_{th}$ to $P_c$ the corrected melting curve is obtained where the shock wave data serve as anchors for the fit (red solid line). This procedure yield the Grüneisen constant $\gamma_o$=2.16(1) nearly in accord with Slater ($\gamma_o$ =2.01(1)) corrected to room temperature [2,13]. Note that within the errors the theoretical calculations of Mariorty et al. [16] (brown star) and Dai et al.[17], corroborate with the present proposed melting curve. The volume compression vs. melting temperature curve is depicted in Fig.1b (blue line) where $\gamma_o$ and $Tm_o$' are those derived above.

In Fig.1c the solid blue line represents the combined approach result in 300K isotherm $B_o,B_o$' 73/4.45. The solid green line is BM 72.7/4.14 obtained by Hänström and Lazor [13]. The magenta solid line represents VIN 73/4.54 proposed by Dewaele at al. [16]. Thus, the experimental data can be well fitted with different EOS and different $B_o,B_o$' parameters. By introducing the present constraint the Vinet 73/4.45 is our best fitting result.

## 2. Copper

A second test of our combined approach is copper metal. Copper at ambient pressure exhibits the fcc structure. Within the resolution of the XRD no phase transitions are observed (Fig.2c) at least up to ~150GPa. The experimental melting data shown in the Fig.2a are taken from Japel et al. [18] and Errandonea [14]. The experimental data points in the P-V plane were measured by Dewaele et al. [19] are shown in Fig.2c. The experimental points in the P-V and P-T are best fitted simultaneously with the LG criterion (eq.5) and Vinet EOS yielding the bulk moduli parameters $B_o$=142 GPa and $B_o$'=4.9 forming $P_c$ and the cold melting curve. The fitting revealed $\gamma_{eff}$= 2.25 where $Tm_o$=1395K is the melting point measured inside the DAC at ambient pressure according to Errandonea (red X in Fig.2a). As shown in Fig.2c the VIN EOS with the constraint imposed on $B_o$ and $B_o$' according to the combined approach perfectly fit the experimental data (300K isotherm). The thermal pressure ($P_{th}$) was calculated according to Altshuler et al. [5] (there in Table IV). Adding $P_{th}$ to $P_c$ according to the procedure 1-2 yield a Grüneisen parameter $\gamma_o$=2.01(3). Here the SW data of Han et al. [20] and Urlin et al. [38] both serve as anchors for deriving the actual pressure melting curve (solid magenta line in Fig.2a). The calculated thermal pressure shift developed in the copper sample confined in the DAC is 8GPa ($Po_{th}$) yielding melting at $Tm_o$'=1656(10)K. In Fig.2b the solid blue line presents the volume compression as function of the melting temperature obtained by the above $\gamma_o$ and $Tm_o$'. The red diamond is shock Hugoniot according to Urlin [38]. The purple O points are derived from shock adiabats measured by Mittchel and Nellis [37] obtained by solving V/Vo for each experimental melting point using LG eq.5.

## 3. α-uranium

α-uranium exhibits orthorhombic (Cmcm) structure and no phase transition was reported up to 300GPa. The experimental melting data points were taken from Yoo,Cynn and Söderlind phase diagram [25]. The experimental data in the P-V plane shown in Fig.3c are derived from references [23,24,25]. The SW data depicted in Fig3a are theoretical calculations of Barroso [28] which is based on first principles and perhaps on experimental measurements. In addition, the theoretical calculation of Dai [29] based on LG criterion (red stars) are also shown in the figure.

The best fits of the experimental (as measured) P-V data which simultaneously fit the melting experimental data are obtained utilizing BM EOS with $B_o$=136(2)GPa and $B_o$'=3.8(0.1) forming $P_c$. These results confirm Yoo,Cynn and Söderlind [26] calculations. The experimental melting points are fitted with the LG criterion eq.5 with $\gamma_{eff}$ = 2.17 and $Tm_o$=1480K forming the cold curve (solid blue line in Fig.3a) marked (136/3.8/2.17) in the figure. Using Forbes [27] simplified approximation with $C_v$= 1.16 erg/gK and $\beta_o$=75 erg/gK$^2$ [28] the thermally corrected pressure $P_{th}$ was calculated. Adding the calculated thermal ($P_{th}$) to $P_c$ the actual pressure melting curve of α-uranium metal is obtained and is depicted in Fig.3a (black solid line). The thermal pressure calculations reveal thermal pressure of 5.45GPa ($Po_{th}$) at zero pressure and melting temperature at $Tm_o$'=1550K. The Grüneisen parameter $\gamma_o$ is obtained by utilizing the procedure 1-2 where the shock wave melting data [28,29] serve as anchor yielding $\gamma_o$ = 2.09.

In Fig.3b depicted is the calculated volume compression as function of the melting temperature obtained by $\gamma_o$ and $Tm_o$' given above. Assuming Vo=20.73 Å$^3$/at. ($\rho_o$=19.0 gr/cm$^3$) the data points (colored stars) are derived by inserting V/Vo in to eq.5 for each melting point. Thus, the blue solid line in Fig.3b presents the volume at the melt. In Fig.3c the EOS in P-V space is shown with all values of $B_o$/$B_o$' suggested by references [23,24,25]. Nevertheless, these suggested bulk moduli fail to fit the experimental melting data no matter which EOS is chosen. For a better insight see discussion and Fig. 5.

## 4. Iron

Under pressure iron metal exhibit several crystallographic phases: α-Fe (bcc), γ-Fe (fcc), ε-Fe (hcp) and β-Fe (dhcp). The melting curve and equation of state of iron ε-phase are depicted in Fig.4a. The green squares are the melting data taken from the phase diagram reported by Anzellini et al. [30] (there in Fig.2). The red diamonds are Murphy's [11] as measured experimental melting points. The magenta and green O's are SW data of Nguyen et al.[31] and Starikov et al.[33] respectively. The black O are the experimental results obtained by Tateno et al. [3]. The black solid line is the melting curve extrapolated to high pressures and temperatures obtained by the present combined approach step 1. Note that the extrapolated melting curve in Fig.4a match the SW data [31,33]. In addition, Tateno's extremely high pressure data indeed indicate that the samples are in the solid phase.

The volume compression vs. the melting temperature is shown in Fig.4b (blue solid line). Solving $V/V_o$ for each melting temperature point reported Anzellini et al.[30], Ma et al.[32], Murphy et al.[11] Mikaylushkin et al.[34], Starikov et al. [33] and Nguyen et al.[31], within the error all are in agreement with the calculated volume vs. melting obtained directly by eq.5. Note that Anzellini at al. (there in Fig.1) indicate that the sample is in the solid phase.

In Fig.4c the red squares are the experimental isotherm 300K EOS according to Dewaele et al. [10]. The solid lines are the VIN, BM EOS (eq.8,9) and the + points (Dubrovinski et al. [40]) all fit well the experimental P-V data. Extrapolation to ambient pressure yield $V_o= 11.15(4)Å^3$. Applying the combined approach step 1 and choosing VIN equation of state reveal the optimized parameters $B_o=163(1)GPa$ and $B'=5.55(0.2)$ with $\gamma_{eff}=1.68(2)$ and $Tm_o=2300K$ (black solid line, Fig.4a). Surprisingly, the present proposed melting curve match the SW data and obviously in contradiction to the isochoric conditions in the DAC claimed above. The reason for this discrepancy will be widely discussed later on. However, as we demand that the SW data must serve as anchor to the melting curve, the black solid line indeed presents the melting curve. In this manner the extrapolation to high pressures and temperatures is legitimate where $\gamma_{eff} = \gamma_o$. We obtain the melting in the inner core boundary melting temperature 5900±100K.

Summery of the derived parameters using the present combined approach is given in Table II:

**Table II.** Derived parameters using the present combined approach: $V_o$ is the volume at ambient temperature and prtessure. $\gamma_{eff}$ is the fitting parameter (replacing $\gamma$ in eq.5) for fitting the experimental measured melting points. $\gamma_o$ is the derived Grüneisen parameter. Assuming isochoric condition in the DAC, the calculated thermal pressure at ambient pressure marked $Po_{th}$. The melting point at $Po_{th}$ is signed $Tm_o$'

| Metal | Phase | $V_o$ Å³/at. | $\gamma_{eff}$ | $\gamma_o$ | $Po_{th}$ GPa | $Tm_o$' °K |
|---|---|---|---|---|---|---|
| Aluminum Al | fcc | 16.6 | 2.45 | 2.16 | 4 | 1150(20) |
| Copper Cu | Cubic fcc | 11.8 | 2.25 | 2.01 | 8 | 1650(20) |
| α -Uranium | Orthorhombic Cmcm | 20.68 | 2.17 | 2.09 | 5.45 | 1550(20) |
| Iron (Fe) | hcp+fcc+dhcp | 11.15 | 1.68 | 1.68 | 25 | 2300(25) |

## 5. Discussion

The prediction of the melting curves at high pressures and temperatures for metals according to the Debye model namely Lindemann-Gilvarry criterion using Grüneisen parameter $\gamma_o$ never fit the measured experimental melting data. The reason is that most previous publications could not decide whether isochoric or isobaric condition exist in the DAC. In the present contribution it is claimed that in any DAC isochoric condition exist no matter the transmitting media, meaning that upon raising the temperature the volume of the measured sample stays constant associated with increase of the melting point. As stated in introduction we apply Lindemamm's criterion for prediction of melting curves thus adopting Lindemann criterion improved by Gilvarry. In many contributions Simon and Glatzel prediction of the melting curve is used [11,30]. This semi empirical formula needs additional two fitting parameters which has no physical meaning contrary to the LG criterion where the fitting parameters are the bulk moduli.

The melting curve of Al metal is shown in Fig.1a (solid red line). The adiabatic volume compression is shown in Fig,1b. The calculated volume compression $V/V_o$ vs. the melting temperatures is depicted by the red solid line in Fig.1b obtained by fixing $\gamma_o$ and $Tm_o$' derived in the V-T plane. Assuming isochoric condition in the DAC the combined approach was performed with $V_o$=10.0 cm$^3$/mol. Aluminum is a good example for demonstrating no discrepancy between the SW data and the DAC experimental results. Indeed, for the first time by performing the correct thermo-physical dynamic calculations a consistent melting curve is achieved with a consistent fitting parameters $B_o$,$B_o$' and where derived $\gamma_o$ is close to Slater prediction [2] at ambient conditions ($\gamma_s$= 2.09, $B_o$'=4.45).

The melting curve of copper at the actual pressure is depicted in Fig.2a (magenta solid line). The combined approach constraints revealed the parameters $B_o$= 142 GPa and $B_o$'=4.9 best fitted with $\gamma_{eff}$= 2.25 forming the cold melting curve (blue line in Fig.2a). Assuming isochoric condition in the heated DAC the melting curve of copper and equation of state (Fig.2c) were calculated according to the procedure in steps 1-2. The calculated melting curve is obtained by introducing actual pressure revealing $\gamma_o$=2.01(3). Within the error, the obtained $\gamma_o$ is in accord with Huang et al. [41,42] hinting that the approximation (q=1) in eq.4 is reasonable. The melting temperature of copper at the actual pressure at ($Po_{th}$) 8Gpa yield $Tm_o$'=1650(20)K.

In Fig.2b depicted is the calculated volume compression as function of the melting temperature obtained by fixing $\gamma_o$ and $Tm_o$'. The experimental data points were derived by solving $V/V_o$ for each melting temperature using LG eq.5. By inserting the shock data $V/V_o$ obtained experimentally by Mitchell and Nellis [37] (assigned as O's in fig.2b) it is clear that shock above 200Gpa ($V/V_o$=0.55-0.58) the liquid state is dominant confirming the observation of V.D. Urline [38]. As shown in Fig.2c isotherm 300K of copper can be best fitted by various models with different $B_o$/$B_o$' parameters. The constraint imposed on the bulk moduli B and B' and the demand that the thermally corrected melting curve should match the SW data (serving as an anchor), indeed indicate that no discrepancy between the SW and the measured experimental data exist. It certainly proves that violation of the isochoric condition in the DAC is of a minor effect. Thus our extrapolation to high pressures and temperatures seems to be reasonable (Fig.1a and Fig.2a).

In Figs.3a,c the melting curve and equation of state of α-uranium are depicted. By demanding that the fit of the experimental melting data should include simultaneously the Lindemann-Gilvarry criterion and the relevant equation of state best fits were obtained by applying BM EOS; $B_o$=136(2) and $B_o$'=3.8(1) and $\gamma_{eff}$ =2.17 (free parameter) with the melting temperature at ambient condition inside the DAC $Tm_o$'=1550K. Indeed, as shown in Fig.3a (blue solid line) a good fit of the experimental melting data is achieved. These results are in agreement with Yoo et al. [26]. We adopt the theoretical calculations of Barroso [28] which is based on first principles and perhaps on experimental measurements as anchor for the α-uranium melting curve. Inserting the thermal correction according to Forbes simplified approximation [27] we find $\gamma_o$=2.09. The calculations of Dai et al. [29] for pressures above 170GPa within the error also confirm these results. The melting temperature of uranium at the calculated actual pressure ($Po_{th}$) 5.45GPa yield $Tm_o$'=1550(20)K. As shown in Fig.5 the BM EOS with the parameters 104/6.2 proposed by Le Bihan et al. [23] or VIN EOS (114.5/5.48) proposed by Dewaele et al. [21] mismatch the SW data, though they fit very well the data in the P-V space (Fig.3c). In conclusion, the uranium case shown in Figs.3a,5 is a good example for demonstrating that the combined approach constraints are indeed reasonable, justifying the extrapolation to high pressures and temperatures.

ε-Iron is perhaps the most interesting case to examine our combined approach . The determination of $B_o$ and $B_o$' under the constraint that they must simultaneously obey the Lindemann-Gilvarry criterion and the relevant EOS, point to VIN EOS revealing **$B_o$=163GPa** and **$B_o$'=5.55**. Indeed, in agreement with the values derived by Dewaele et al.[10]. The melting curve was obtained with $\gamma_{eff}$ =1.68 forming $P_c$. Moreover, as shown in Fig.4a, the present combined approach with these parameters fit very well the experimental melting data derived by Anzellini [30] and within the error Murphy's [13] melting data. But, most astonishing result is that the DAC experimental results and the SW experimental data show no discrepancy between the two techniques. The combined approach steps 1-2 demands are that the melting SW data should serve as anchor for deducing the melting curve while taking in to account the actual pressure, meaning that in the iron case $\gamma_{eff} = \gamma_o$=1.68 (Fig.4a). In the following we try to speculate on the possible reasons of this phenomenon:

The phase diagram of iron above 100GPa was measured in a heated DAC by Anzellini et al. indicating that the ε phase is dominant up to 200GPa ([30], there Fig.2). In addition, in-situ XRD measurements (there Fig. 1c) shows volume increase namely isobaric condition in the DAC. This result is in absolute contradiction to our claim that in the DAC isochoric condition exit. We speculate that the origin of the volume expansion observed while raising the temperature and pressure relate to the fact that approaching the melt a mixture of the phases $\varepsilon,\beta,\gamma$ exist. Thus isochoric condition in the DAC should not be denied. In the following we will try to clarify our speculation:

The existence of the γ phase at high pressures and temperatures have been reported by Mikaeylushkin et al. [30] showing that the γ phase can exist even at 165 GPa when quenched to room temperature (RT), namely no triple point. In addition, S.K. Saxena al et. [39] reported observation of the β phase above 110GPa and 3000K. Deep look in Fig.1a of Anzellini et al. at 133GPa and 4292K (green dots) at the angle 11.58(2Θ) there exist a possible reflection which is ignored by the authors. This reflection could be analyzed as (111) gamma (γ) phase or dhcp (100) (β) phase. Thus, while approaching the melt a possible mixture of $\varepsilon,\beta,\gamma$ phases could lead to the observed discrepancy. The observation of the volume increase is by no means connected to the DAC or to the transmitting medium. As shown in Figs.4a,b different experiments with deferent DACs and different pressure transmitting media (Anzellini, Murphy, Mikhaylushkin and Ma) show the same volume increase. To conclude, the volume expansion claimed by Anzellini at al. is a solid fact that deserve further studies.

**Conclusions**

Assuming isochoric conditions in the DAC the Lindemann-Gilvarry criterion is applicable for predicting the melting curves of Al,Cu and U. By introducing the constraint demanding that the fitting of the experimental EOS (P-V space) data will simultaneously fit the experimental melting results and demanding that the shock wave melting data (P-T space) will serve as anchor for the fit, the discrepancies between the shock wave and the DAC data are settled. This lead to direct determination of the Grüneisen parameter $\gamma_0$. Thus, a safe extrapolation of the melting curve to high pressures and temperatures is achieved. It can not be denied that volume expansion observed while raising the temperature and pressure above 100GPa observed in iron metal, relate to the fact that approaching the melt a mixture of the phases $\varepsilon,\beta,\gamma$ exist.

# Figures and captions

Fig.1

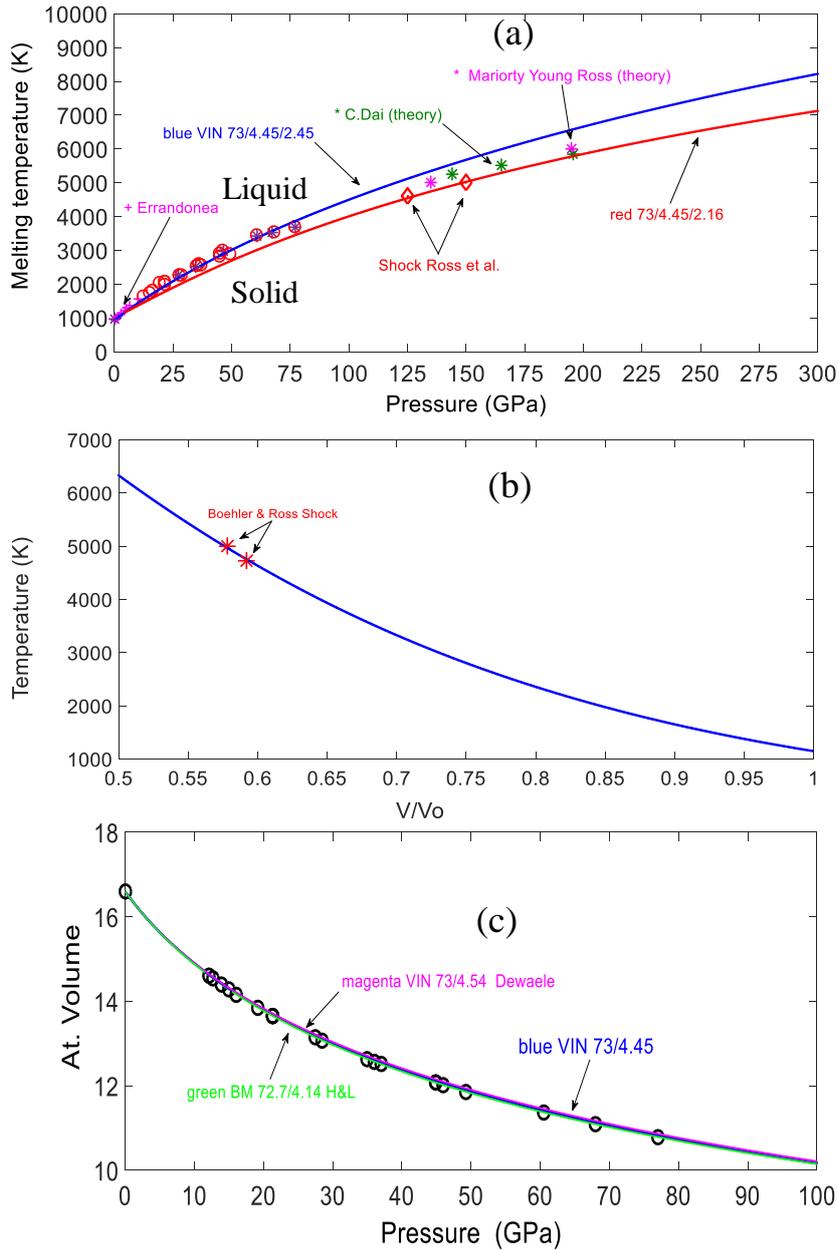

**Fig.1** : The melting curve of Aluminum is presented. The blue solid line represents the fitting of the as measured experimental data using Lindeman-Givarry melting criterion with $B_o$=73(0.5)GPa and $B_o$'=4.45 and $\gamma_{eff}$= 2.45 (assigned B/B'/$\gamma$). Assuming isochoric condition in the heated DAC the melting curve is obtained by introducing the thermally corrected pressure (actual pressure), revealing $\gamma_o$= 2.16 and $Tm_o$'=1150K (red solid line). In Fig.1b depicted is the calculated volume compression as function of the melting temperature. In Fig.1c the EOS isotherm 300K best fitted with the above bulk moduli parameters $B_o, B_o'$. Note that the experimental EOS data can be best and different EOS and different $B_o, B_o'$ fit parameters (see text) .

Fig.2

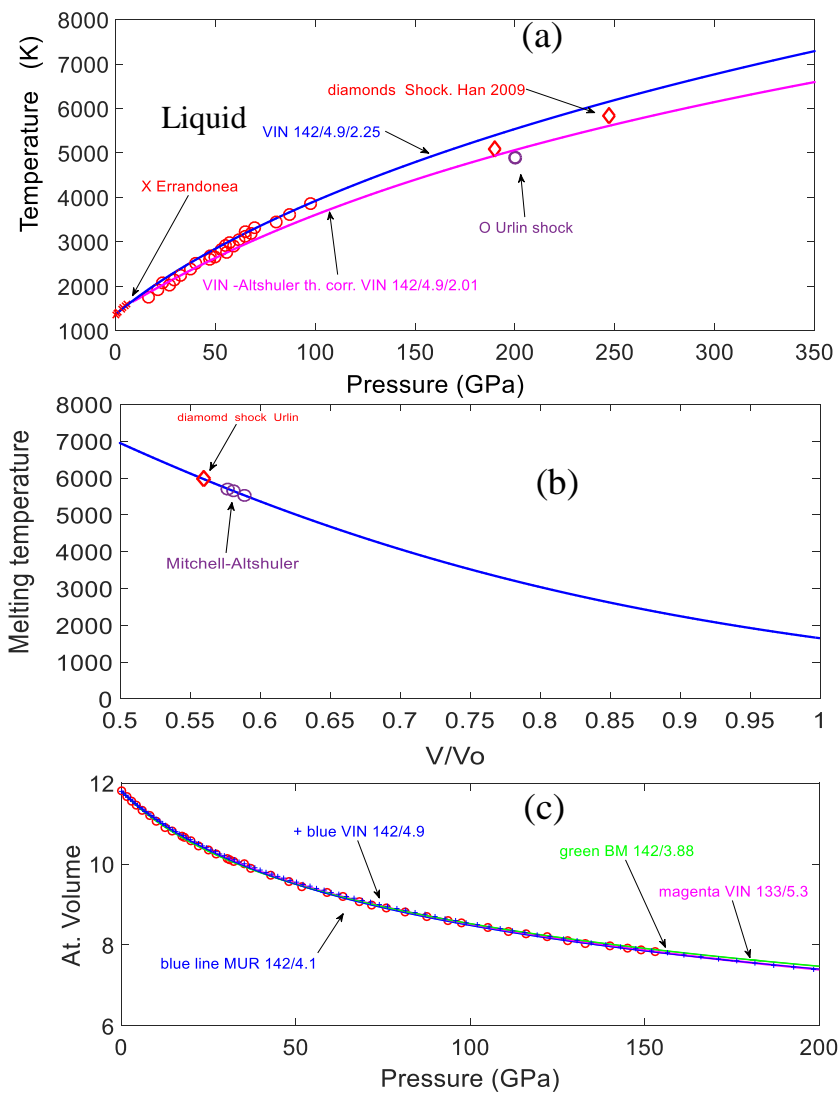

**Fig.2 :** a: The melting curve of copper metal is presented. The blue solid line represents the fitting of the experimental (as measured) data with $B_o=142(2)$GPa and $B_o'=4.9$ using the combined approach (step 1) where $\gamma_{eff}= 2.25$ (assigned $B/B'/\gamma$) and $Tm_o=1395$K. The magenta solid line is the melting curve as function of the actual pressure (thermally pressure corrected) assuming isochoric condition in the heated DAC revealing $\gamma_o= 2.01(3)$ and $Tm=1650(20)$K. In Fig.2b depicted is the volume compression as function of the melting temperature. The red diamond is shock is shock Hugoniot according to Urlin [38]. The purple O points are derived from shock Hugoniot of Mitchel and Nellis [37] obtained by solving V/Vo for each experimental point using LG eq.5. In Fig.2c the EOS isotherm 300K (blue line) best fitted with the above bulk moduli. Note that the experimental EOS data can be best fitted by various models and different $B_o, B_o'$ parameters.

Fig.3

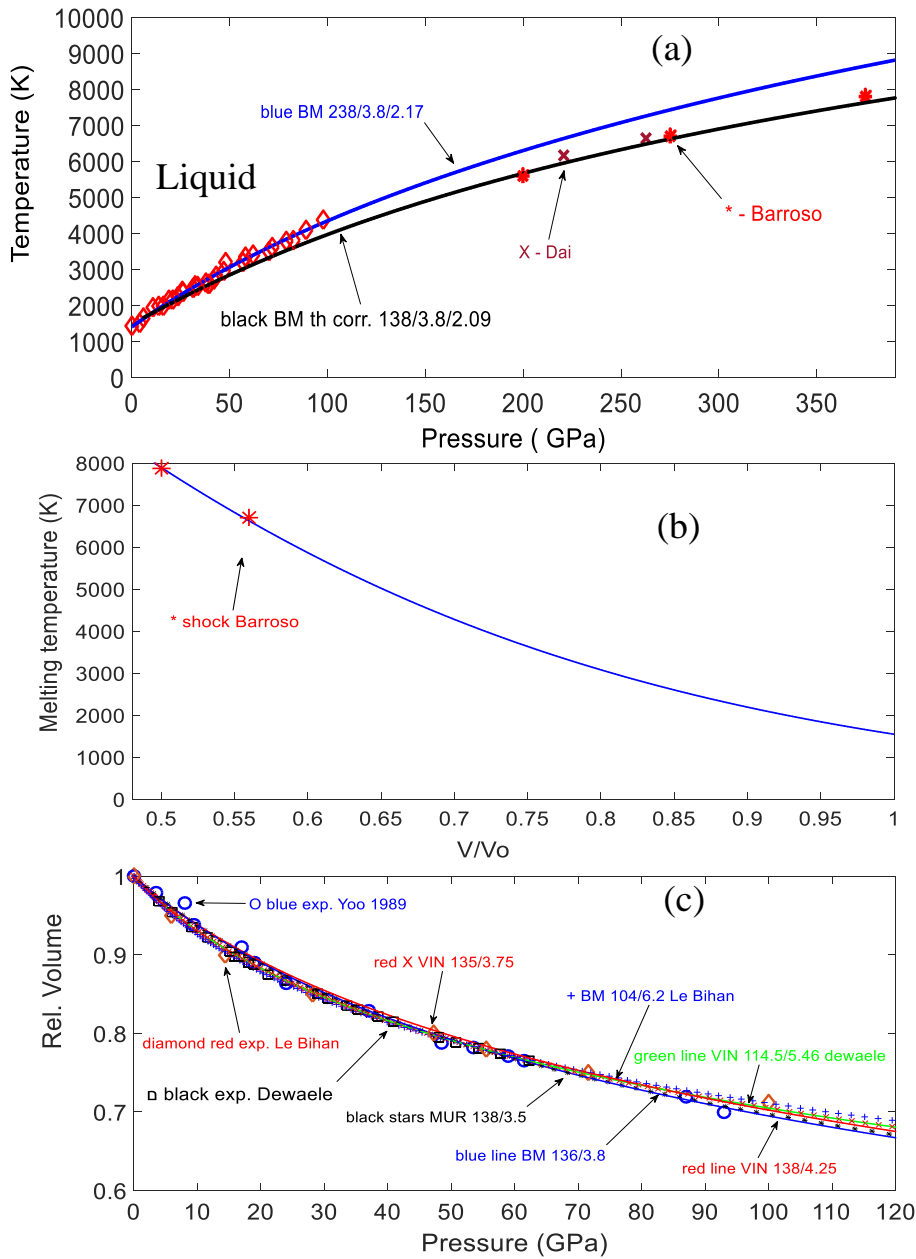

**Fig.3 : a**: Melting curve of α- uranium vs. pressure is presented by the black solid line. The solid blue line is the fit of the EOS which simultaneously fit the experimental uranium melting (as measured) data (combined approach). The fitting procedure revealed $B_o$= 136(2) GPa and B'=3.8(1) and $\gamma_{eff}$ = 2.17 with $Tm_o$=1500K (assigned B/B'/γ) forming $P_{cold}$. The solid lines are obtained using the BM EOS combined with Lindeman melting criterion. The red stars and brown X in are the shock wave theoretical values [17,28]. The black solid line represents the melting curve vs. actual pressure derived the Grüneisen parameter $\gamma_o$ = 2.09 and $Tm_o$'=1550K. **b**: Depicted is the calculated volume compression as function of the melting temperature. **c**: EOS in P-V space. The blue O are experimental data are taken from ref. [25]. The red diamonds and black squares come from ref. [23,24] respectively. Note that all values of $B_o/B_o$' fit the EOS experimental data no matter which EOS is chosen.

Fig.4

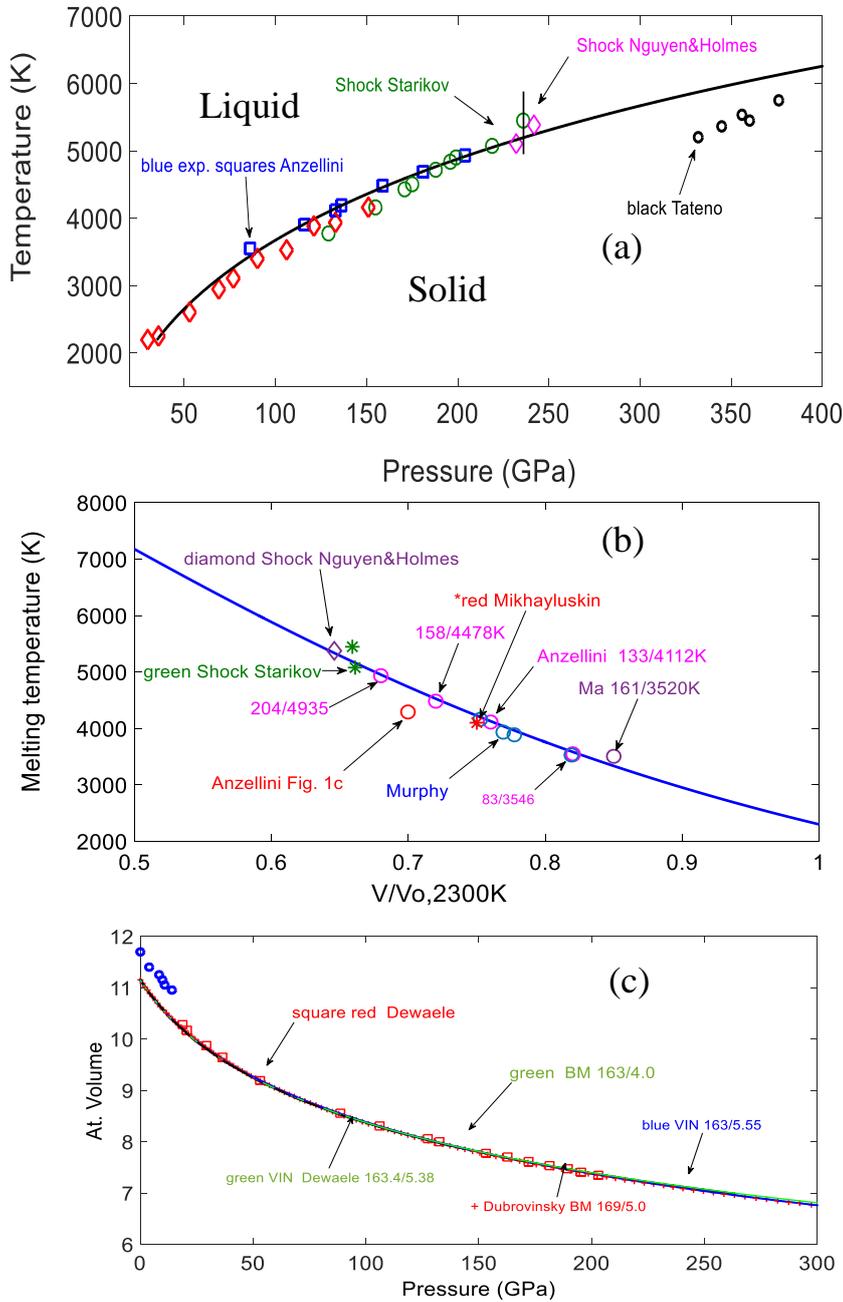

**Fig.4** : (a): Melting curve and quation of state of iron metal are depicted in Fig.4a. The blue squares are the melting data reported by Anzellini et al. phase diagram [29] and the red diamonds are C.Murphy et al. [13] experimental meting data. The black O are the experimental results obtained by Tateno et al. [5]. The black solid line in (a) is the combined approach extrapolation to high pressures and temperatures based on melting data of Anzellini's (blue squares) and Murphy's (red diamonds). (b): Pressure vs. volume at the melt by solving eq.5 blue solid line. The experimental points were derived from Anzellini et al. [30], Ma at al.[32], Starikov etal.[33], Murphy et al.[13], and Mikhayloshkin et al.[34] by solving V/Vo out from eq.5. explained in the text. In (c) the red squares are the experimental equation of state (EOS) according to Dewaele et al. [29]. The solid blue line is the VIN (eq.8) EOS (isotherm 300K) which simultaneously best fit the experimental P-V data and the melting curve data according to the procedure 1-4 combined approach. Extrapolation to ambient pressure give $V_{o,300K}$ = 11.15 Å$^3$/at. Finally, our calculations reveal

$B_o$=163(1)GPa and B'=5.55(0.2). However, as shown in (c) the experimental data can be fitted with different bulk moduli parameters.

Fig. 5

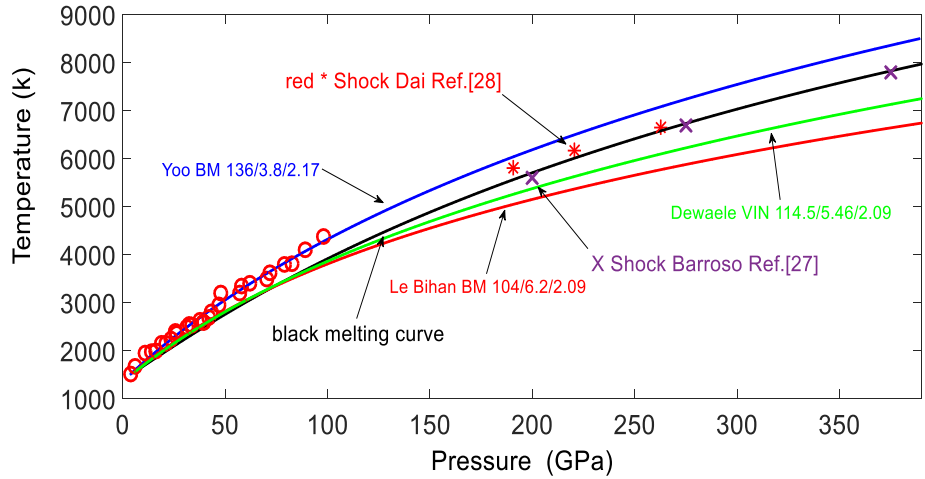

**Fig.5 :** Melting curve of α- uranium vs. pressure is presented. By utilizing the combined approach (steps 1-2) demanding that the fit of the EOS (P-V) will simultaneously fit the P-T thermally corrected melting curve where the SW data should serve as anchor, the BM with the parameters 136/3.8 yield the best simultaneous fits. However, with the bulk moduli parameters 104/6.2 and 114/5.46 proposed by Le Bihan et al. [23] and Dewaele at al. [24] the melting curve miss the SW data (green and red solid lines).